\documentclass[journal]{IEEEtran}
\usepackage{hyperref}
\hypersetup{colorlinks,allcolors=black}

\usepackage{bm}
\usepackage{mathrsfs}
\usepackage{amsmath}
\usepackage{amssymb}
\usepackage{graphicx}
\usepackage{subfigure}
\usepackage{stfloats}
\usepackage{float}
\usepackage{multirow}
\usepackage{cite}
\usepackage{color}
\ifCLASSINFOpdf
\else
\fi
\begin{document}
\title{An Adaptive MMC Synchronous Stability Control Method Based on Local PMU measurements}

\author{Long Peng,~\IEEEmembership{Student Member,~IEEE}, Yong Tang,~\IEEEmembership{Senior Member,~IEEE}, Lamine Mili, \IEEEmembership{Life Fellow,~IEEE}, Yingbiao Li, Bing Zhao,~\IEEEmembership{Member,~IEEE}, Yijun Xu,~\IEEEmembership{Member,~IEEE}, Fan Cheng
	
\thanks{This work was supported by National Key Research and Development Program of China (2016YFB0900600) and Technology Projects of State Grid Corporation of China (52094017000W).}

\thanks{L. Peng, Y. Tang, Y. Li, B. Zhao and F. Cheng  are with China Electrical Power Research Institute. (e-mail: penglong\_epri@163.com; tangyong@epri.sgcc.com).}

\thanks{L. Mili and Y. Xu is with the Bradley Department of Electrical and Computer Engineering, Virginia Polytechnic Institute and State University, Falls Church, VA 22043 USA (e-mail: lmili@vt.edu; yijunxu@vt.edu).}}

\markboth{IEEE TRANSACTIONS ON POWER SYSTEMS, 2020}%
{Shell \MakeLowercase{\textit{et al.}}: Bare Demo of IEEEtran.cls for IEEE Journals}

\maketitle

\begin{abstract}
Reducing the current is a common method to ensure the synchronous stability of a modular multilevel converter (MMC) when there is a short-circuit fault at its AC side. However, the uncertainty of the fault location of the AC system leads to a significant difference in the maximum allowable stable operating current during the fault. This paper proposes an adaptive MMC fault-current control method using local phasor measurement unit (PMU) measurements. Based on the estimated Thevenin equivalent (TE) parameters of the system, the current can be directly calculated to ensure the maximum output power of the MMC during the fault. This control method does not rely on off-line simulation and adapts itself to various fault conditions. The effective measurements are firstly selected by the voltage threshold and parameter constraints, which allow us to handle the error due to the change on the system-side. The proposed TE estimation method can fast track the change of the system impedance without depending on the initial value and can deal with the TE potential changes after a large disturbance. The simulation shows that the TE estimation can accurately track the TE parameters after the fault, and the current control instruction during an MMC fault can ensure the maximum output power of the MMC.

\end{abstract}

\IEEEpeerreviewmaketitle
\section{Introduction}
\IEEEPARstart
{S}{}ince the advent of high-voltage direct current (HVDC) transmission systems based on the voltage source converter (VSC), they are playing an increasingly important role in power systems \cite{tang2010voltage,flourentzou2009vsc}.
The application of modular multilevel converter (MMC) greatly improves the VSC-HVDC transmission capacity \cite{yingbiao2017analysis}. Besides, the  characteristics of the MMC also have a great impact on the transient characteristics of a power system.
The existing studies on MMC mainly focus on small signal stability \cite{ni2017improved} and the DC  fault characteristics \cite{ZHAOXibei5724}, whereas the  synchronization stability caused by the AC side fault is rarely studied. In \cite{wu2018adaptive,taul2019efficient,wu2019design}, it was found that, when there is a short-circuit fault in the AC power system, the MMC is at risk of a loss of synchronization (LOS) with the AC grid. The synchronization stability of the MMC is closely related to the system's Thevenin equivalent (TE) potential and impedance and to the MCC phase-locked loop (PLL) parameters and output active power current.
Reducing the active current of the MMC after the fault can effectively improve the synchronization stability. Furthermore, the amplitude of the current regulation is closely related to the equivalent potential and impedance of the system after the fault.
However, due to the uncertainty of the fault location in the AC system, the difference between the equivalent potential and the inductance is large.
In \cite{wu2018adaptive}, the MCC active power current is reduced to zero, which greatly improves the synchronization stability of the MMC during the fault. However, it has a negative impact on the stability of the AC power system by reducing the MCC active power to zero. To increase the active power of the MMC, the regulation of the active power current should be adjusted according to the TE equivalent potential and impedance of the system after the fault. Then, the maximum power transmission is realized and the transient characteristics of the AC power system can be improved while ensuring its synchronous operation with the MMC. It can be seen that the estimation of the TE parameters of the system after failure becomes the key to control the MCC output current and change it to a desired value.

The TE voltage and impedance depend on the power system topology and parameters \cite{yun2019online}. However, the TE of a large-scale system is difficult to accurately evaluate in real-time. Facing this challenge, methods based on local measurements have been proposed in the literature \cite{vu1999use}. They  are mainly applied in static voltage stability, which requires a change of the voltage and current measurements on the load-side so as to guarantee system observability. Some of the existing methods rely on accurate initial values  \cite{wang2012real,babazadeh2017real}, otherwise the measurements are required to be sufficient enough to track the parameters. Moreover, it is generally assumed that the TE parameters are constant \cite{zhao2016robust}, or at least that the amplitude of the TE potential is constant \cite{abdelkader2014online}. If this is not the case, then alternative methods should be employed. For instance, Kalman filter methods may be utilized to estimate the TE parameters\cite{hoffmann2012online}, but they rely on the preset covariance matrices of the process and measurement noise. Furthermore, sufficient effective measurements are also required to ensure the convergence. 

Regarding the MMC connected to an AC power system, it has the following characteristics. First, the MMC has a fast response, and there will be a large voltage and current change at the beginning of the fault, but it can quickly realize the output current tracking reference current and reach the new steady state. When the voltage and current enter the new steady state, the system parameters are not observable. Second, the fault is usually accompanied by the sudden change of TE parameters, so the initial value of parameters after the failure is unknown.
Thirdly, TE impedance is generally kept constant after failure, but due to the dynamic response of the system side, TE potential will continuously change.
In addition to the above problems, in order to meet the needs of MMC stability control after failure, it is also required to track parameter changes within tens of milliseconds, and the lack of valid measurements also makes it difficult for some recursive estimation methods to ensure accurate identification\cite{wang2012real,babazadeh2017real,hoffmann2012online}.
Therefore, it is difficult to realize fast parameter tracking under large disturbance in the existing TE parameter estimation methods.

To address the aforementioned issues, an adaptive control method to enhance MMC  synchronous stability during failure is proposed. According to the voltage change and parameter constraints, the effective measurements from the local PMU are first selected to reduce the influence of the system-side disturbance. Then, the least squares method is used to estimate the TE parameters. Finally, according to the synchronous stability control characteristics of MMC, the current control instruction is obtained to realize the maximum power transmission under the synchronous and stable operation of MMC during the fault.

The rest of the paper is organized as follows. {{In Section \uppercase\expandafter{\romannumeral2}, the synchronous stability of MMC connected to an ac system are analyzed. In Section \uppercase\expandafter{\romannumeral3}, the TE estimation method  is proposed to track the change of the impedance. Numerical results are conducted and analyzed in Section \uppercase\expandafter{\romannumeral4}. Finally, Section \uppercase\expandafter{\romannumeral6} concludes the paper.}}

\section{Synchronous stability analysis of MMC}

\subsection{System Description}
\begin{figure}
\setlength{\abovecaptionskip}{-0.1cm} 
\centering
\includegraphics[scale=0.4]{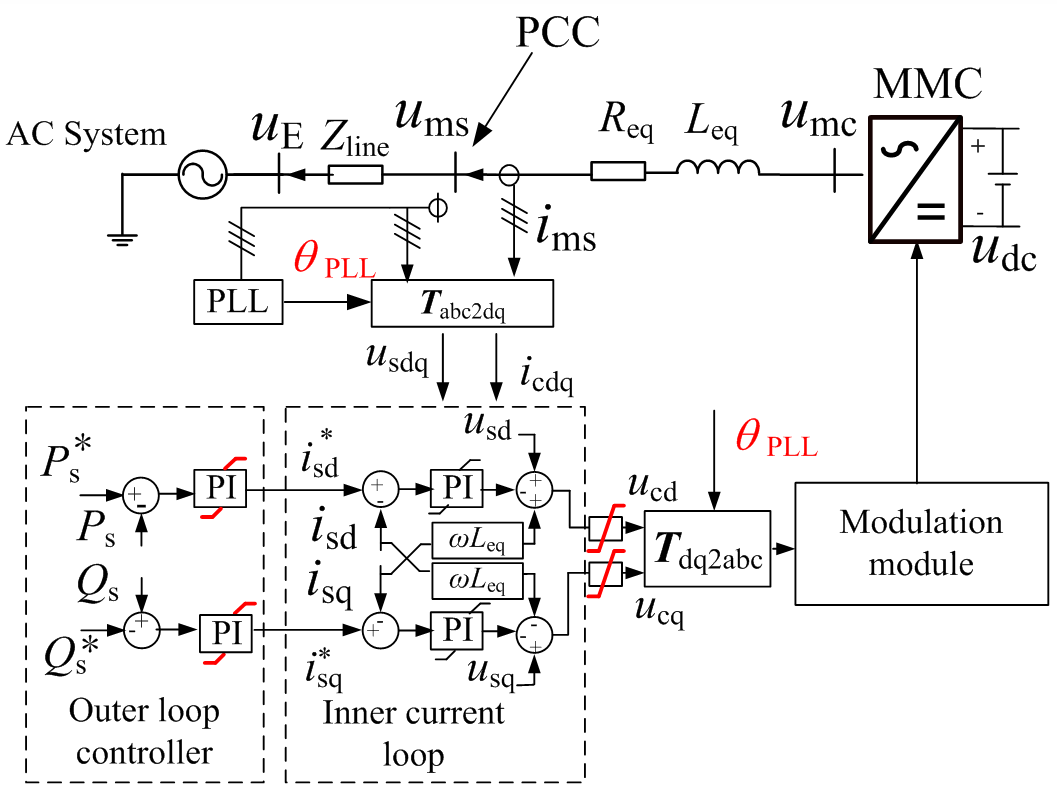}
\caption{Diagram of MMC with control loops}
\label{1}
\end{figure}

Fig. \ref{1} shows the equivalent topology structure of MMC connected to ac power grid. MMC adopts the control mode of constant power control. $u_{ms}$ denotes the voltage at the point of common connection (PCC), $u_{mc}$ is the output voltage of MMC. $i_{ms}$ is MMC output current, $P_s$ and $Q_s$ are the active and reactive power output of MMC, $u_e$ denotes equivalent voltage of ac power grid, $Z_{line}$ denotes system equivalent impedance, the $R_{eq}$ and $L_{eq}$ are MMC equivalent resistance and inductance, $\theta_{PLL}$ is PLL output phase, reference superscript "*" denotes  references, the subscript "dq" denotes $dq$ coordinates component . $T$ represents parker change and the transformation function. Based on PLL, MMC can detect the frequency and phase of voltage of connected point, and then  the active power and reactive power output by MMC can be controlled by controlling the output voltage of $u_{mc}$. Therefore, the normal operation of PLL is the basis for MMC to keep synchronously with ac system.
\begin{figure}[htb]
\setlength{\abovecaptionskip}{-0.1cm} 
	\centering
	\includegraphics[scale=0.35]{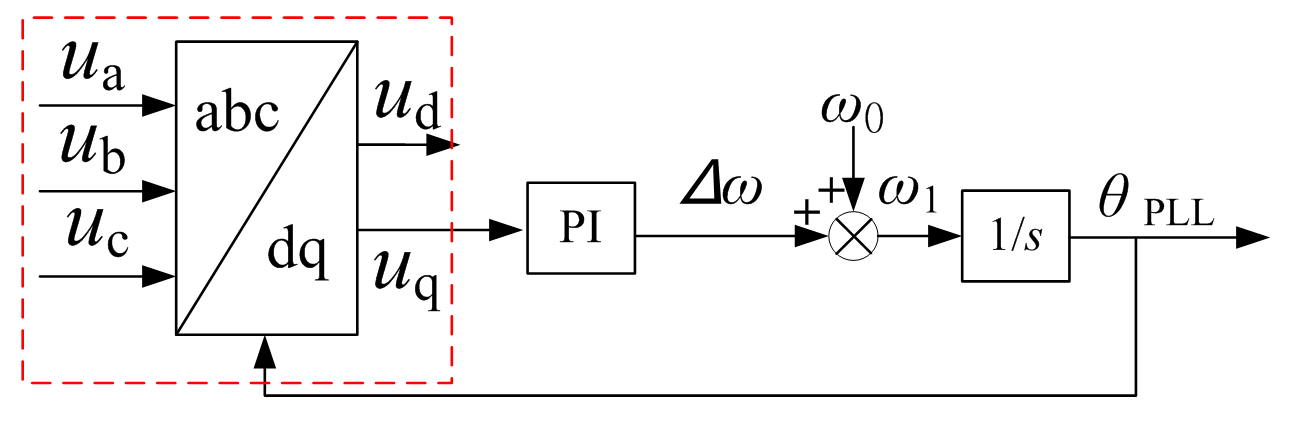}
	\caption{The control of PLL}
	\label{2}
\end{figure}
\begin{figure}[htb]
\setlength{\abovecaptionskip}{-0.1cm} 
	\centering
	\includegraphics[scale=0.25]{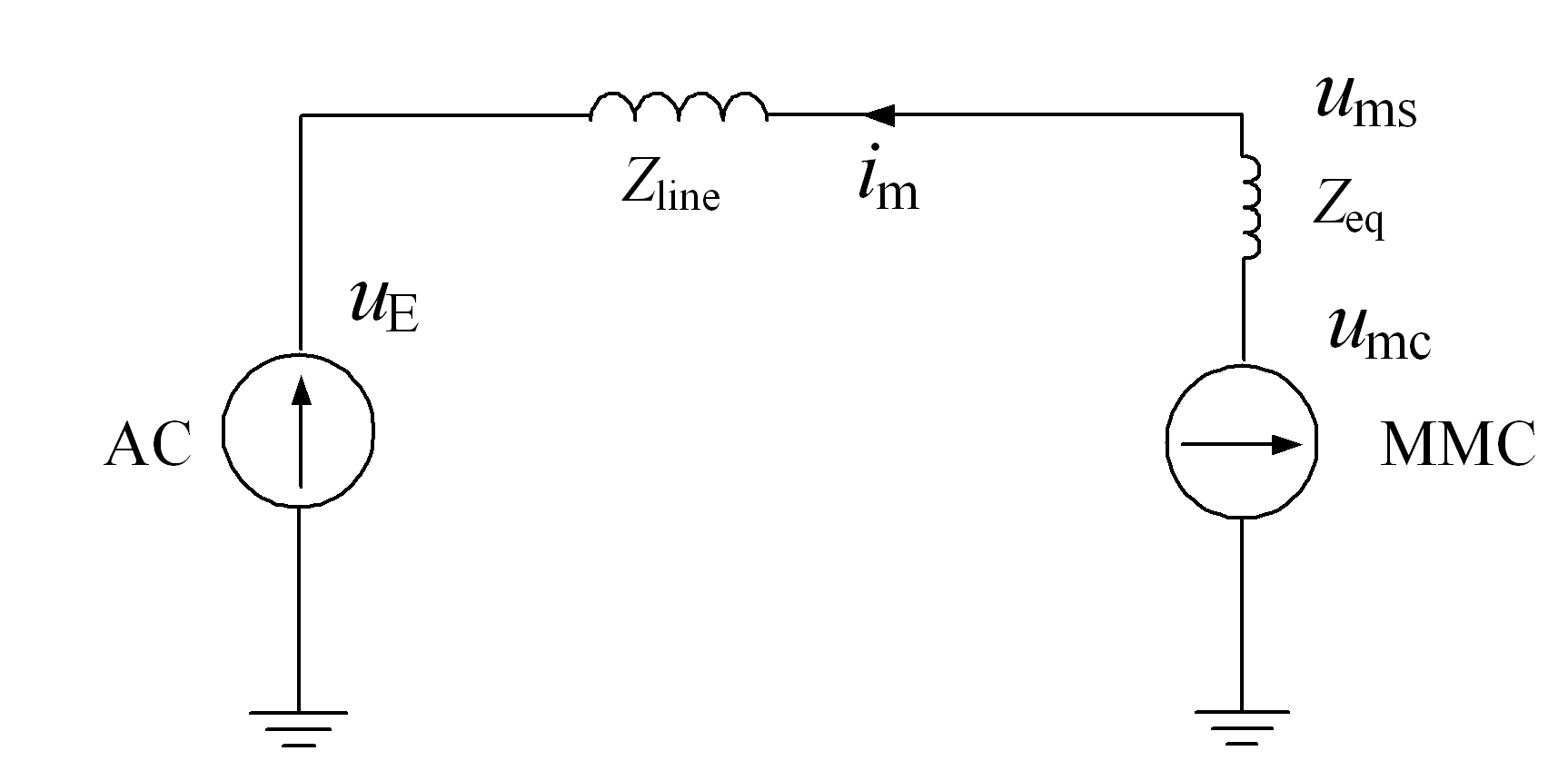}
	\caption{Equivalent circuit}
	\label{3}
\end{figure}
\subsection{Existence of Equilibrium Points }
The control block diagram of PLL is shown in Fig. \ref{2}. Firstly, the three-phase voltage is transformed to the dq coordinate system through the park transformation, and the q-axis voltage is input into the PI controller, so as to track the voltage phase and frequency of the system. In Fig. \ref{2}, $\omega_{0}$ is the fundamental frequency of the system, $\omega_{1}$ is the output frequency of PLL, and $\theta_{PLL}$ is the output phase angle of PLL. Similarly, $\theta_{PLL}$ provides the angle  for the MMC park transformation and valve control. From Fig. \ref{2}, the condition for the existence of equilibrium points of PLL can be defined as $u_{msq}$  = 0. When a three-phase short circuit fault occurs in the ac system, MMC can be equivalent to the current source \cite{guangquan2017analysis}, and the AC system  can be equivalent to the structure of voltage source and impedance in series. Its equivalent circuit is shown in Fig. \ref{3} and the voltage of q-axis at PCC point can be solved as
\begin{equation}
   {{u}_{\text{msq}}}={{u}_{\text{Eq}}}+{{\omega }_{1}}{{L}_{\text{line}}}{{i}_{\text{md}}}+{{i}_{\text{mq}}}{{R}_{\text{line}}}
  \label{be1} 
\end{equation}
In the dq coordinate system of MMC, $u_{Eq}$ can be expressed as
\begin{equation}
{{u}_{\text{Eq}}}=-{{u}_{\text{E}}}~\sin \left( {{\theta }_{\text{PLL}}}-{{\theta }_{\text{E}}} \right)=-{{u}_{\text{E}}}\sin \delta
  \label{be2} 
\end{equation}
where ${{\theta }_{E}}$ is the phase of the system equivalent voltage source, $\delta $  is the phase difference between the MMC and the system equivalent voltage source. Consequently, (\ref{be1}) can be expressed as 
\begin{equation}
{{u}_{\text{msq}}}=-{{u}_{\text{E}}}\sin \delta +{{\omega }_{1}}{{L}_{\text{line}}}{{i}_{\text{md}}}+{{i}_{\text{mq}}}{{R}_{\text{line}}}
  \label{be3} 
\end{equation}
In order to ensure the existence of a steady-state operating point of the system, PLL needs to be able to operate normally, that is, $u_{msq}$ = 0 has a solution, which can be known
\begin{equation}
{{\omega }_{1}}{{L}_{\text{line}}}{{i}_{\text{md}}}+{{i}_{\text{mq}}}{{R}_{\text{line}}}={{u}_{\text{E}}}\sin \delta \le {{u}_{\text{E}}}
  \label{be4} 
\end{equation}
It can be seen from (\ref{be2}) that in order to ensure the existence of a steady-state operating point of the system, the output current of MMC needs to be adjusted according to the system equivalent potential and impedance after the fault.
\subsection{Transient Characteristic Analysis}
\begin{figure}[H]
\setlength{\abovecaptionskip}{-0.1cm} 
	\centering
	\includegraphics[scale=0.22] {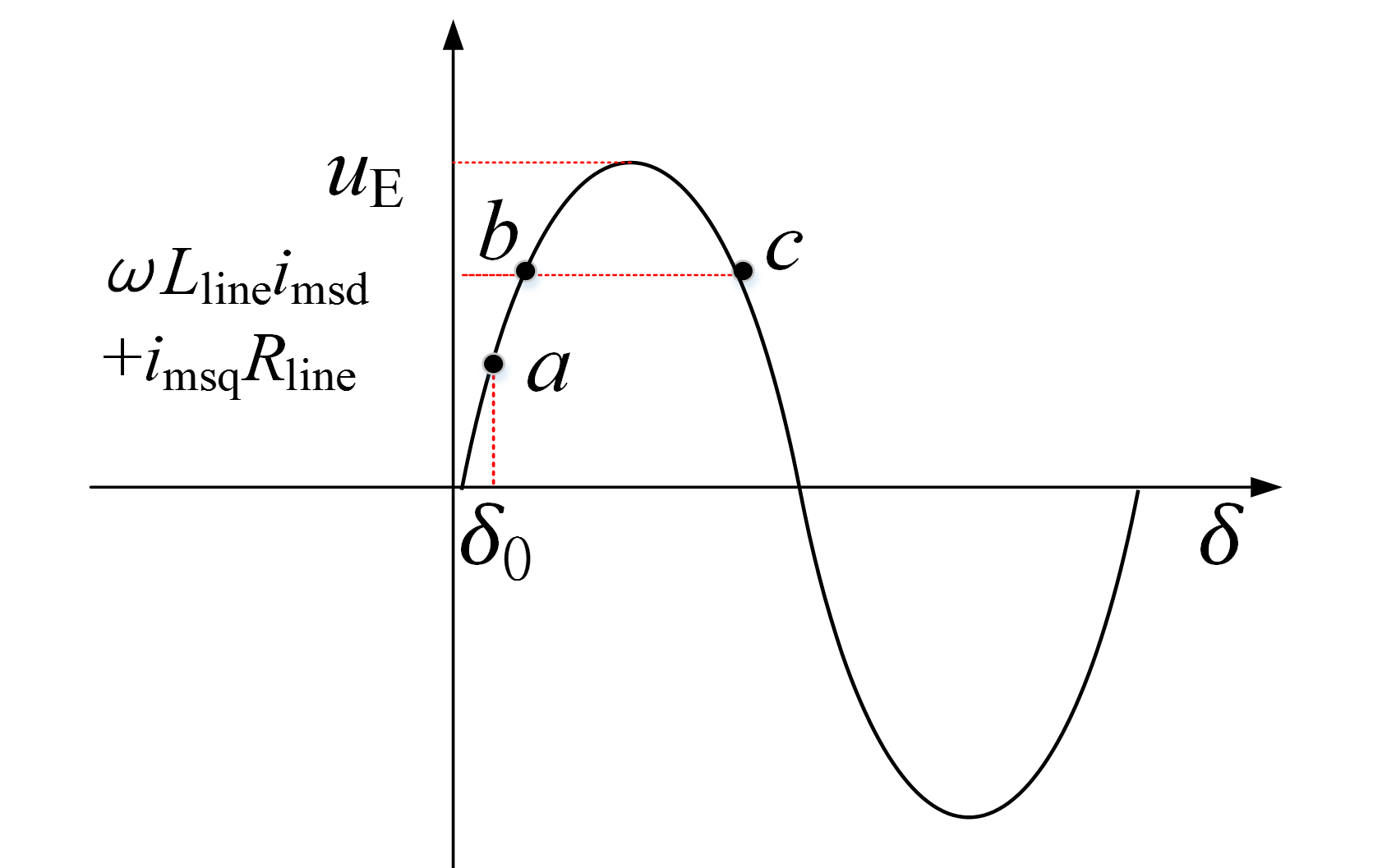}
	\caption{Diagram of equilibrium}
	\label{4}
\end{figure}

According to the PLL control block diagram shown in Fig. \ref{2}, it can be obtained that the phase angle of the output of the phase-locked loop is

\begin{equation}
{{\theta }_{\text{PLL}}}=\int{{{\omega }_{1}}\operatorname{d}t}
  \label{be5} 
\end{equation}
where
\begin{equation}
\begin{aligned}
  & {{\omega }_{1}}={{\omega }_{0}}+\Delta \omega  \\ 
 & \Delta \omega ={{K}_{p\text{PLL}}}\cdot {{u}_{\text{msq}}}+{{K}_{i\text{PLL}}}\cdot \int{{{u}_{\text{msq}}}\text{d}t} \\ 
 \end{aligned}.
\label{be6}
\end{equation}
where, ${{K}_{p\text{PLL}}}$ and ${{K}_{i\text{PLL}}}$ are the proportional coefficient and integral coefficient of PLL proportional integral (PI) controller respectively. Since
\begin{equation}
{{\theta }_{\text{E}}}=\int{{{\omega }_{0}}\operatorname{d}t},
  \label{be7} 
\end{equation}
the phase angle between MMC and the ac system can be expressed as

\begin{equation}
\delta ={{\theta }_{\text{PLL}}}-{{\theta }_{\text{E}}}=\int{\left( {{K}_{\text{pPLL}}}\cdot {{u}_{\text{msq}}}+{{K}_{\text{iPLL}}}\cdot \int{{{u}_{\text{msq}}}\text{d}t} \right)\text{ d}t}
\label{be8}
\end{equation}
When there are steady-state operating points in the system, it is shown in Fig. \ref{4}. Set the initial power angle as $\delta_{0}$, then the initial operating point of the system after the fault is point a. According to (\ref{be3}), when $u_{msq} > 0$, then $\Delta \omega > 0$, the output frequency of PLL at MMC side will be higher than the frequency $f_0$ on the system side. Therefore, the output phase angle of MMC will increase, and the system operating point  moves  from point a to point b. When arriving at point b, $u_{msq}$ = 0, but due to the integrator of the PI controller, $\Delta \omega$ is still higher than 0. As a result, the frequency of MMC is still higher than the system frequency, so the operating point exceeds point b to point c. That means $u_{msq}$ is going to be less than 0 after it crosses b. Before reaching the point c, if $\Delta \omega < 0$, output phase angle of PLL  will decrease, and begin to return to point b, with the final stable at point b. When reaching point c, if $\Delta \omega$ is still higher than 0, the frequency of MMC is higher than the system frequency. After exceeding the point c, $u_{msq} > 0$, the output phase angle of MMC is further enlarged, causing the MMC lose of synchronization with the ac system. So point b is the steady-state operating point, and point c is the unstable operating point. According to Fig. \ref{4}, when there is a steady-state operating point in the system, the output current of MMC has an affect on the position of MMC's steady-state operating point and the synchronization stability of MMC. Therefore, it is still necessary to further adjust the current output of MMC according to the system equivalent potential and impedance after the failure.

Substituting (\ref{be3}) into (\ref{be8}), the power angle equation of MMC and ac system can be solved as

\begin{equation}
\begin{aligned}
& \overset{\centerdot \centerdot }{\mathop{\delta }}= \frac{{{K}_{\text{iPLL}}}\left[ {{i}_{\text{msd}}}\left( {{\omega }_{0}}+\overset{\centerdot }{\mathop{\delta }} \right){{L}_{\text{line}}}+{{i}_{\text{msq}}}{{R}_{\text{line}}}-{{u}_{\text{E}}}\sin \delta  \right]}{1-{{K}_{\text{pPLL}}}{{i}_{\text{msd}}}{{L}_{\text{line}}}} \\
&-\frac{\overset{\centerdot }{\mathop{\delta }}{{K}_{\text{iPLL}}}{{u}_{\text{E}}}\cos \delta }{1-{{K}_{\text{PPLL}}}{{i}_{\text{msd}}}{{L}_{\text{line}}}}.\\
\end{aligned}
\label{be9}
\end{equation}

Generally, the damping coefficient of the PLL is set to 0.707 \cite{WU2019Analytical}, that is, the PI parameter of the phase-locked loop is certain, and the system's equivalent potential and impedance are certain after the fault of the ac system. Therefore, (\ref{be9}) can be used for quantitative analysis of the influence of d-axis current. 
\begin{figure}[H]
\setlength{\abovecaptionskip}{-0.1cm} 
	\centering
	\includegraphics[scale=0.25] {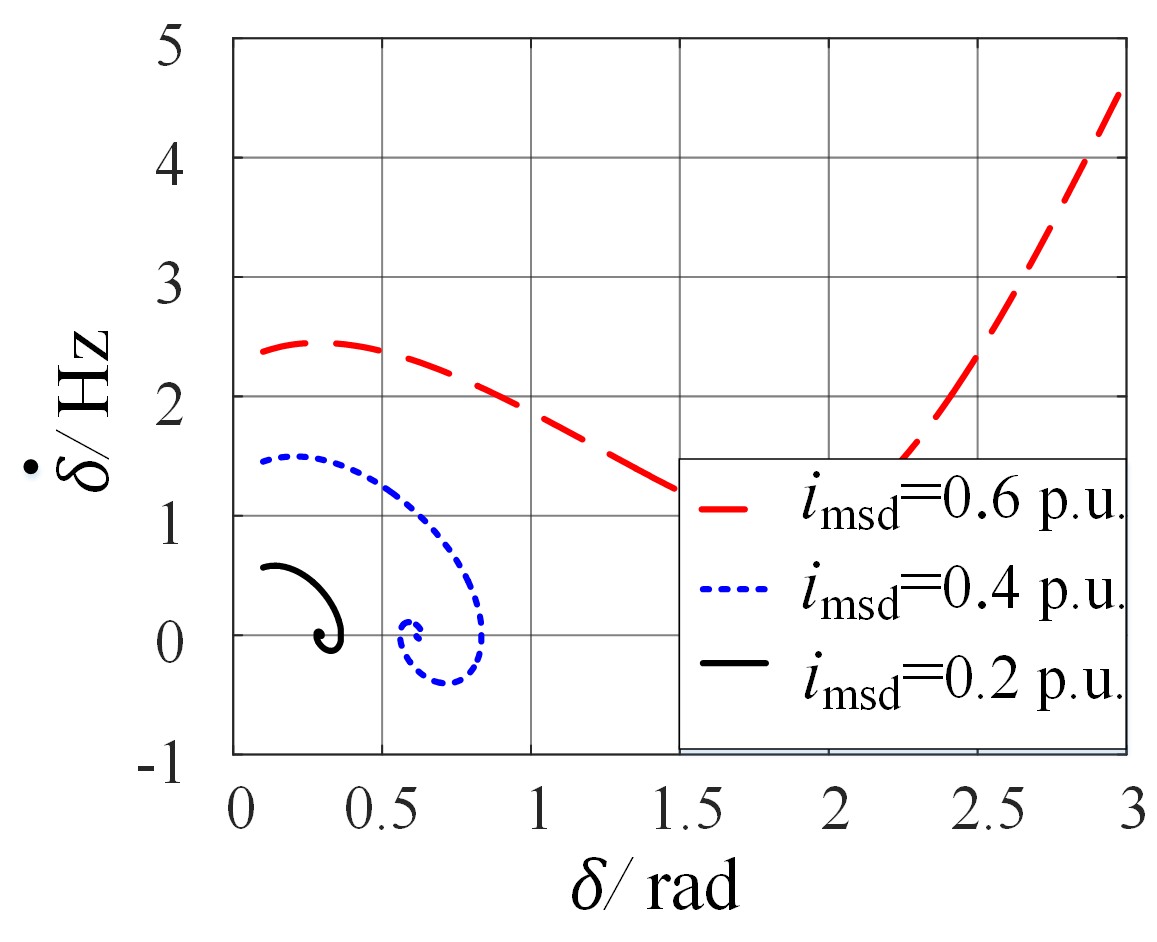}
	\caption{Phase portraits in different $i_{msd}$}
	\label{5}
\end{figure}

The effect of MMC output current on transient stability is shown in Fig. \ref{5}. It can be seen that when the d-axis current is $0.8 \, \text{pu}$, the power angle cannot converge and the MMC loses its synchronization stability with AC system. When the d-axis current is $0.6\, \text{pu}$ or $ 0.4\, \text{pu}$, the power angle can converg, and the MMC can keep synchronous with the system. It can be seen that when the system equivalent potential and equivalent impedance are constant, the MMC output d-axis current has an upper limit $I_{max}$, which can be obtained through (\ref{be9}). In addition, due to the small resistance of the system, the influence of the current on the transient stability of the q-axis can be ignored. In order to ensure the synchronization stability of the MMC, the d-axis current  during the fault should not exceed $I_{max}$, but in order to improve the power support during the fault, the MMC should try to increase the d-axis current. Therefore, the fast detection of the equivalent potential and equivalent reactance after system fault is of great significance to the control of MMC.

\section{TE Parameter Estimation of a Power System Subject to a Large Disturbance}
After a fault, the TE impedance will change due to the change of the system topology, so only post-fault measurements are valid for estimating the updated impedance. In general, the impedance can remain unchanged at least for a short period due to the discrete characteristic of the topology change. Thus, assuming the impedance is constant can still hold if only using the post-fault measurements. However, the TE potential will change due to the dynamic response on the system-side. To this end, the influence of the system-side change on the impedance estimation should be considered.

\subsection{Error Analysis of the Impedance Estimate}
\begin{figure}[htb]
\setlength{\abovecaptionskip}{-0.1cm} 
	\centering
	\includegraphics[scale=0.6] {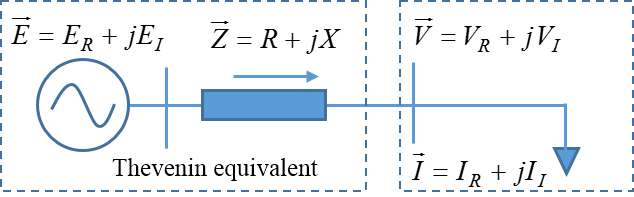}
	\caption{TE seen from the load bus}
	\label{6}
\end{figure}

First, let us display the TE equivalent circuit of the AC system  in Fig. \ref{6}. Within it,  $\vec{E}$ is the TE potential, $\vec{Z}$ is the TE impedance, $\vec{V}$ is the bus voltage and $\vec{I}$ is the current. 

Based on the Kirchhoff's law, the following equation holds:
\small
\begin{equation}
\left( \begin{matrix}
{{I}_{R}} & -{{I}_{I}} \\
{{I}_{I}}  & {{I}_{R}}\\
\end{matrix} \right)\left( \begin{matrix}
R  \\
X  \\
\end{matrix} \right)=\left( \begin{matrix}
{{V}_{R}}  \\
{{V}_{I}}  \\
\end{matrix} \right)-\left( \begin{matrix}
{{E}_{R}}  \\
{{E}_{I}}  \\
\end{matrix} \right).
\label{e1}
\end{equation}
\normalsize

For two continuous PMU samples, the following equation can be obtained:
\begin{equation}
\left( \begin{matrix}
{{dI}_{R}} & -{{dI}_{I}} \\
{{dI}_{I}}  & {{dI}_{R}}\\
\end{matrix} \right)
\left( \begin{matrix}
R  \\
X  \\
\end{matrix} \right)=\left( \begin{matrix}
d{{V}_{R}}  \\
d{{V}_{I}}  \\
\end{matrix} \right)-\left( \begin{matrix}
d{{E}_{R}}  \\
d{{E}_{I}}  \\
\end{matrix} \right),
\label{e3}
\end{equation}
\normalsize
where $d$ represents the variations within these two samples. It can be further simplified as
\begin{equation}
   dI\cdot Z=dV-dE, 
  \label{e4} 
\end{equation}
yielding 
 
 \begin{equation}
\frac{1}{\left\| d{{I}^{-1}} \right\|\left\| dV-dE \right\|}\le \frac{1}{\left\| Z \right\|}\le \frac{\left\| dI \right\|}{\left\| dV-dE \right\|},
\label{e5}
 \end{equation}
where $\left\|\cdot \right\|$ represents the Euclidean norm. Then we get 
 \begin{equation}
       \frac{1}{\left\| Z \right\|}\le \frac{\left\| dI \right\|}{\left\| dV-dE \right\|}.
\label{e6}
 \end{equation}
Now, we can reformulate (\ref{e4}) as 
 \begin{equation}
  dI\cdot \left( Z+dZ \right)=dV.
  \label{e7}
\end{equation}
Here, $dZ$ denotes the impedance error caused by $dE$, from which we infer that $dZ={dI}^{-1}dE$. Thus, we have
 \begin{equation}
\frac{\left\| dE \right\|}{\left\| dI \right\|}\le \left\| dZ \right\|\le \left\| d{{I}^{-1}} \right\|\left\| dE \right\|.
  \label{e8}
\end{equation}
Define that the relative error $\delta ={\left\| dZ \right\|}/{\left\| Z \right\|} $. Using (\ref{e6}) and (\ref{e8}),  we have 
\begin{equation}
 \frac{1}{\left\| dI \right\|\left\| d{{I}^{-1}} \right\|}\frac{\left\| dE \right\|}{\left\| dV-dE \right\|}\le \frac{\left\| dZ \right\|}{\left\| Z \right\|}\le \left\| d{{I}^{-1}} \right\|\left\| dI \right\|\frac{\left\| dE \right\|}{\left\| dV-dE \right\|}.  
\label{e9}
\end{equation}
 Since $d{{I}^{-1}}=d{{I}^{T}}/ \text{det}(dI)$, it can be concluded that
 \begin{equation}
  \|d{{I}}\|=\sqrt{\max \operatorname{eig}\left[(d{I})^{T}d{I}\right]}=\sqrt{ \text{det}(dI)}
\label{e10},
 \end{equation}  
 
 \begin{equation}
\|d{{I}^{-1}}\|=\sqrt{\max \operatorname{eig}\left[({d{{I}^{-1}}})^{T} d{I^{-1}}\right]}=\sqrt{\frac{1}{ \text{det}(dI)}},    
\label{e11}
 \end{equation} 
yielding ${\left\| d{{I}^{-1}} \right\|} \cdot {\left\| dI \right\|}=1$. Consequently, (\ref{e9}) can be simplified as 
\begin{equation}
  \delta =\frac{\left\| dZ \right\|}{\left\| Z \right\|}= \frac{\left\| dE \right\|}{\left\| dV-dE \right\|}.  
\label{e12}
 \end{equation}
When $\left\| dV \right\|\ge \left\| dE \right\|$,
\begin{equation}
\delta=\frac{\left\| dE \right\|}{\left\| dV-dE \right\|}\le \frac{\left\| dE \right\|}{\left\| dV \right\|-\left\| dE \right\|}=\frac{1}{\left\| dV \right\|/\left\| dE \right\|-1}.  
\label{e133}
 \end{equation}
Consequently, if it holds that
\begin{equation}
 {\|dV\|} \ge (\frac{1}{k\%}+1)\|dE\|
\label{e14},
 \end{equation} 
then $\delta$ will be less than $k\%$. If we can find the upper bound of $\|dE\|$ , we can choose the measurements that satisfies 
 \begin{equation}
{\|dV\|} \ge (\frac{1}{k\%}+1){dE}_{{\text{max}}}
\label{e35}
\end{equation}
to achieve accurate estimation.

\subsection{The Upper Bound of the TE Potential Change} 
Obviously, the dynamic voltage response in the power system will affect the  abovementioned ${dE}$. Note that the TE potential is the load-side open-circuit voltage. 
Based on the equivalent circuit reduced to the dynamic regulators, such as the generators, we have
\begin{equation}
\vec{E}(\vec{{Y}_{0}}+\sum\limits_{i=1}^{m}{ \vec{{Y}_{i}}})-\sum\limits_{i=1}^{m}{\vec{{{E}_{Gi}}}} \vec{{Y}_{i}} =0 ,
\end{equation}
where $ \vec{{E}_{Gi}} $ is the regulator potential, $m$ is the number of regulators, $\vec{{Y}_{i}} = G_i+jB_i$ is the branch admittance, $\vec{{Y}_{0}} = G_0+jB_0$ is the  transfer admittance to ground on the load-side. By neglecting the transfer conductivity $G_i$ of the branch and the susceptance $B_0$ to ground, we get the following incremental form:
\begin{equation}
\vec{dE}({G}_{0}+\sum\limits_{i=1}^{m}{ {j{B}_{i}}})
=\sum\limits_{i=1}^{m}{ \vec{{{dE}_{Gi}}}} {{jB}_{i}}, 
\end{equation}
yielding
\begin{equation}
\bigl| \vec{dE} \bigr|=\frac{\bigl| \sum\limits_{i=1}^{m}{ \vec {dE}_{Gi} }  {jB}_{i} \bigr|} 
{\bigl|{G}_{0}+\sum\limits_{i=1}^{m}{  j{B}_{i} }\bigr|} 
\leq \frac{\sum\limits_{i=1}^{m}{ \bigl| \vec{ {dE}_{Gi} }  }  \bigr| \bigl| { {B}_{i} } \bigr|} 
{   \sum\limits_{i=1}^{m}{   \bigl|{B}_{i} \bigr|   }    }\leq \max\left\{{\bigl| \vec{ {dE}_{Gi} }  }  \bigr|\right\} .
\end{equation}
Note that $\bigl| \vec{dE} \bigr|$ is less than the variation of the fastest regulator in the vicinity of the MMC. So ${dE}_{\text{max}}$ can be set to $\max\left\{{\bigl| \vec{ {dE}_{Gi} } }  \bigr|\right\}$, which is the deviation between the initial value and its maximum regulating limitation. In the latter, the excitation system of the generator is taken as an example to analyze ${dE}_{\text{max}}$. More specifically, for a generator using  a flux-decay model that is connected to an infinite bus, whose voltage is described as $U\angle 0$ connected to a series impedance $X_S$, its internal potential ${{E}'}_{q}$ is expressed as
\begin{equation}
 {T_{d0'}}\frac{{d{{E'}_q}}}{{dt}} + {{E'}_q} = {E_{fd}}-(X_d-{X'}_d)I_d,
\label{k29}
\end{equation}
\begin{equation}
{E'}_{q}-({{X'}_{d}}+{{X}_{S}}){{I}_{d}}-{{U}}\cos{{\delta }} =0,
\label{k30}
\end{equation}
where $T_{d0'}$ is the d-axis transient open-circuit time constant, $E_{fd}$ is the no-load potential, $X_d$ is the d-axis synchronous reactance, ${X'}_d$ is the d-axis transient reactance, $I_d$ is the d-axis current, and $\delta$ is the rotor angle. The incremental form can be written as

\begin{equation}
{{T}_{d{0}'}}\frac{d\Delta {{{{E}'}}_{q}}}{dt}+\Delta {{{{E}'}}_{q}}=\Delta {{E}_{fd}}-(X_d-{X'}_d)\Delta{I_d},
\label{k15}
\end{equation}
 
\begin{equation}
\Delta{E'}_{q}-({{X'}_{d}}+{{X}_{S}})\Delta{{I}_{d}}=0, 
\label{k16}
\end{equation}
where the change of $\delta$ is neglected thanks to the fast sample rate of the PMU device. Then we have

\begin{equation}
{{T}_{d{0}'}}\frac{d\Delta {{{{E}'}}_{q}}}{dt}+C\Delta {{{{E}'}}_{q}}=\Delta {{E}_{fd}},
\label{k17}
 \end{equation}
where $C=(X_d+X_S)/({X'}_d+X_S)$, yielding

\begin{equation}
  \frac{d\Delta {{{{E}'}}_{q}}}{dt}=\frac{\Delta {{E}_{fd}}}{{{T}_{d{0}'}}}{{e}^{-\frac{C}{{{T}_{d{0}'}}}t}}\le \frac{\Delta {{E}_{fd{\text{max}}}}}{{{T}_{d{0}'}}}{{e}^{-\frac{C}{{{T}_{d{0}'}}}t}} \le \frac{\Delta {{E}_{fd{\text{max}}}}}{{{T}_{d{0}'}}},
\label{e23}
 \end{equation}
where $ \Delta {E_{fd \text{max}}} $ is the maximum change of the no-load potential. In order to reduce the estimation error, $||dE||$ should be as small as possible. To this end, we choose the minimum sampling time $\Delta T$. Then the upper bound of the TE potential change in the two adjacent measurements is given by 
\begin{equation}
  {dE}_{\text{max}}= \frac{\Delta {{E}_{fd{\text{max}}}\Delta T}}{{{T}_{d{0}'}}}.
\label{e24}
 \end{equation}
 
 Considering typical parameters ${{{T}'}_{d0}}$ = 5s, $\Delta {{E}_{fd\text{max}}}$ = $5\, \text{pu}$ \cite{kundur1994power}, $\Delta T$= 0.01s, then $d{{E}_{\max }}$ = $0.01 \, \text{pu}$. Substituting $d{{E}_{\max }}$ into (23), it can be known that if the error less than 10 $\%$ is required, the screening threshold of voltage should be $0.11\, \text{pu}$.
\subsection{TE Estimation Considering the Parameters Constraints} 
Note that the upper bound in (\ref{e24}) may be too large for some minor faults. From (\ref{e23}), we can see that $dE$ at two adjacent measurements will decrease with time. Hence, we can reduce the threshold appropriately to improve the applicability of the method to minor faults. To reduce the uncertainties brought by the adjustment of the threshold, we propose to add some constraints for the parameters. The reasonable constraints for the impedance and potential can be set as 
\begin{equation}
  0 \le X \le X_{\text{max}}, R \le X,  \bigl| \vec{E} \bigr| \le 1 \, \text{pu}.
\label{e25}
\end{equation}
The constraints above are explained as follows. Since the three-phase fault is equivalent to a new grounding branch connected in parallel with the network topology, the equivalent impedance of the system will decrease, and $X_{\text{max}}$ can be set as the reactance calculated off-line before the fault. In addition, the failure is generally accompanied by the decrease of equivalent potential, so the upper bound of the potential can be set as $1 \, \text{pu}$. In addition, for high voltage power grids, $R$ is usually less than $X$. Above all, with the constraints of TE parameters taken into account, this paper recommends the voltage threshold of ${\|dV\|} \ge 0.02 \, \text{pu}$.

\subsection{Main Steps of TE Estimation Algorithm} 
Now, we summarize the main steps of the proposed method as shown below.

Step 1. Selecting the measurements. We select the measurements within window $m$ after the disturbance. The measurements should meet the two following conditions:

{\romannumeral1}. ${\|dV\|} \ge 0.02 \, \text{pu}$. 

{\romannumeral2}. The TE parameters calculated by (\ref{e1}) and (\ref{e3}) assuming $dE=0$ are bounded by (\ref{e25}).
Otherwise, the impedance will not be updated. 

Step 2. Estimating the impedance.  At time $n$, the impedance ${\hat{Z}}_n$ can be solved using the selected measurements by
 \small
 \begin{equation}
  {\hat{Z}}_n= ({{H_n}^{T}H_n})^{-1}{H_n}^{T}{Y}_n, 
  \label{e26} 
\end{equation}
where 
\begin{equation}
H_n=\left( \begin{matrix}
{{{dI}}_{n-m+1}}  \\
{{{dI}}_{n-m+2}}  \\
\vdots   \\
{{{dI}}_{n}}  \\
\end{matrix} \right), 
Y_n=\left( \begin{matrix}
{{{dV}}_{n-m+1}}  \\
{{{dV}}_{n-m+2}}  \\
\vdots   \\
{{{dV}}_{n}}  \\
\end{matrix} \right).
\end{equation}
\normalsize

Here, $dI_n$ and $dV_n$ represents $dI$ and $dV$ in time $n$, respectively.
 
\section{Numerical Results}
This section utilizes two cases to verify the proposed method. Firstly, an ideal voltage source system is adopted, and secondly, MMC is considered to be connected to a two-area system with dynamic response characteristics.
\subsection{Ideal Voltage Source System}
\begin{figure}[H]
\setlength{\abovecaptionskip}{-0.1cm} 
	\centering
	\includegraphics[scale=0.25] {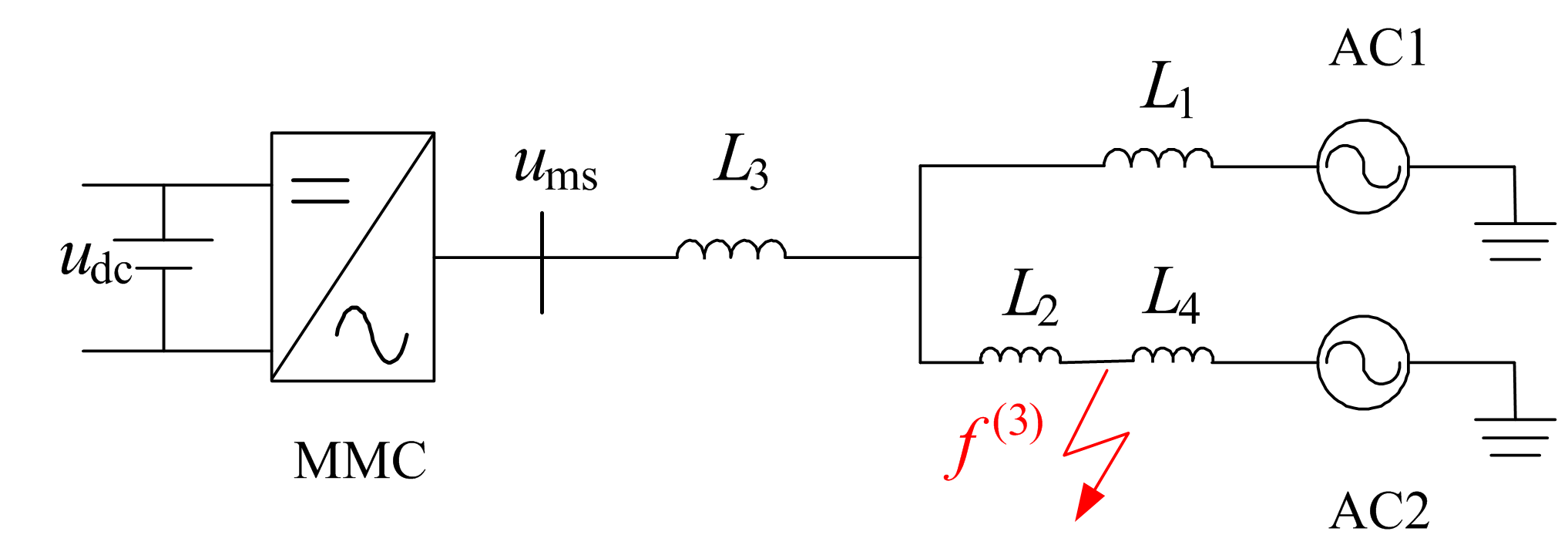}
	\caption{System topology under ideal voltage source}
	\label{7}
\end{figure}

The topology under the ideal voltage source is shown in Fig. \ref{7}. The output power of MMC is set to be $0.8\, \text{pu}$. At 0.5 s, a three-phase short circuit fault occurred at the position shown in Fig. \ref{7}. At 1.1 s,the $L_2$ branch is disconnected and the fault is cleared. The system parameters are as follows: ac rated voltage 230 kV, dc rated voltage 500 kV, MMC rated power 750 MW, operating power 600MW. PLL damping coefficient 0.707, PLL limiting $\pm2.5\,\text{Hz}$ , $L_1 = 0.05\, \text{H}$, $L_2= 0.02\, \text{H}$, $L_3= 0.1\, \text{H}$, $L_4= 0.01\, \text{H}$, voltage source potential of $1\, \text{pu}$.

\subsubsection{TE parameter estimation}
\begin{figure}[H]
\setlength{\abovecaptionskip}{-0.1cm} 
	\centering
	\includegraphics[scale=0.48]{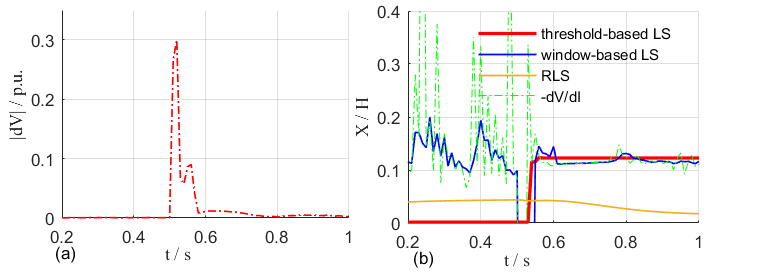}
	\caption{TE estimation considering ideal voltage source. (a) $||dV||$ when subject to a large disturbance; (b) the comparison of TE parameters estimation using different methods}
	\label{8}
\end{figure}

according to the circuit, the equivalent inductance of the system during the failure is calculated to be 0.107 H, and the equivalent voltage $0.285\, \text{pu}$. 
This paper utilize three common methods as comparison. They are,respectively, window-based least squares (LS), recursive least squares (RLS) and $-dV/dI$ which is based on the ratio of voltage change and current change at adjacent moments. The voltage variation $||dV||$ and estimated inductance are shown in Fig. \ref{8}. Using the proposed threshold-based LS method, the measurements larger than the voltage threshold and meeting the constraints are selected for estimation. After the failure, the equivalent inductance of the system is identified as 0.115 $H$ at 0.54 s. The TE potential was calculated as $0.27\, \text{pu}$ by (\ref{e10}). RLS method cannot obtain the accurate estimation. This is mainly because there is not a correct initial value after the failure. For window-based LS and -dV/dI methods, the inductance identified at the initial stage of the fault has a large oscillation, and the inductance can be approximately tracked slowly than the proposed method. Meanwhile, the estimated impedance is not stable, resulting in a large fluctuation. Above all, the proposed method based on the threshold shows the advantage to the inductance estimation.
\begin{figure}[H]
\setlength{\abovecaptionskip}{-0.1cm} 
	\centering
	\includegraphics[scale=0.21] {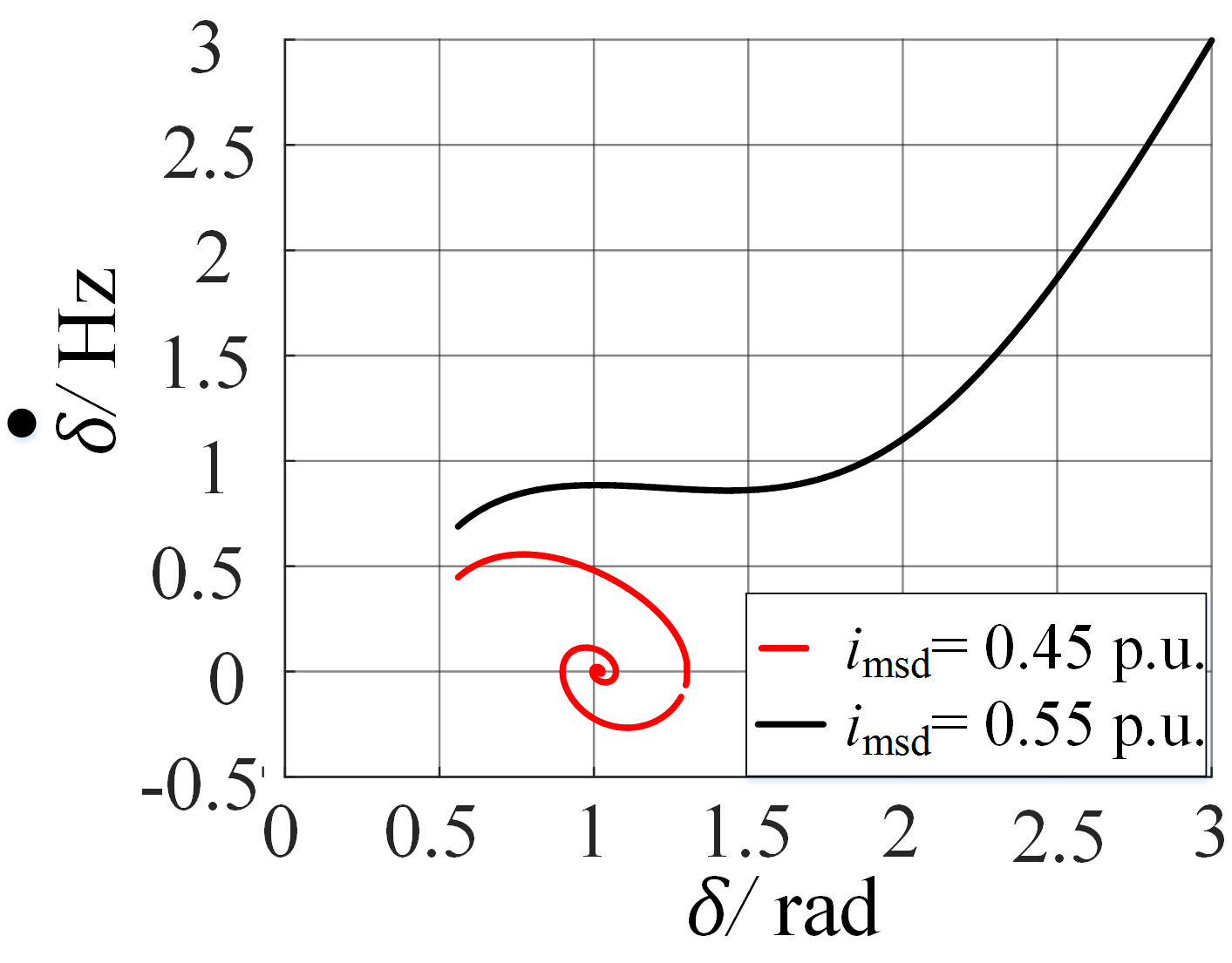}
	\caption{Phase portraits of MMC based on TE estimation}
	\label{9}
\end{figure}
\subsubsection{Output current control of MMC}
\begin{figure}[H]
\setlength{\abovecaptionskip}{-0.1cm}   
	\centering
	\includegraphics[scale=0.5]{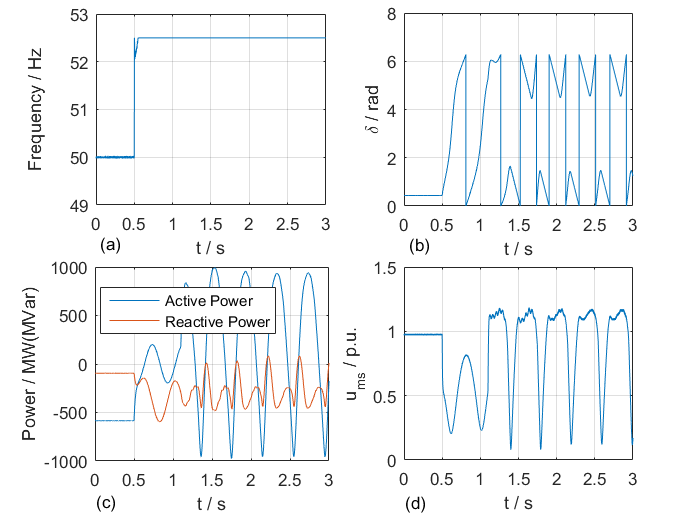}
	\caption{{{Simulation curves when $i_{sd}$ = $0.55 \, \text{pu}$. (a) frequency; (b) angle; (c) active/reactive power (d) voltage.}}}
	\label{10}
\end{figure}

\begin{figure}[H]
\setlength{\abovecaptionskip}{-0.1cm}   
	\centering
	\includegraphics[scale=0.5]{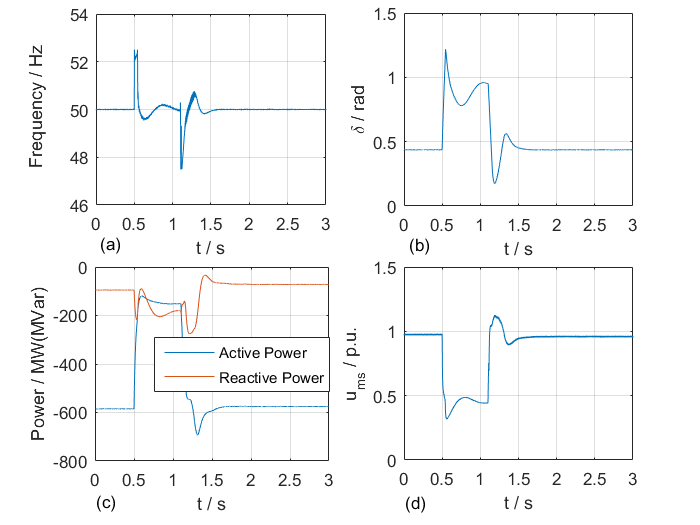}
	\caption{{{Simulation curves when $i_{sd}$ = $0.45 \, \text{pu}$. (a) frequency; (b) angle; (c) active/reactive power (d) voltage.}}}
	\label{11}
\end{figure}

substituting the estimated inductance into (\ref{e9}), then, when the output d-axis current of MMC is larger than $0.45\,\text{pu}$, MMC cannot keep in sync with the system. Therefore, the output current of MMC should be less than $0.45\,\text{pu}$, and the curve at different current levels is shown in Fig. \ref{9}. The simulation results are shown as Fig. \ref{10}, where $i_{sd}$ is controlled as $0.55\,\text{pu}$ during the short circuit fault. It can be seen that, the output frequency of PLL reaches the limit value and is higher than the voltage source frequency. Therefore, the MMC is asynchronous with the system as shown in Fig. \ref{10}(b). Meanwhile, it can be seen that, from Fig. \ref{10}(c) and (d), when  MMC is asynchronous with the system, the voltage and the power are out of control and unstable. However, when $i_{sd}$  is controlled as $0.45\,\text{pu}$ during the  fault, the MMC can keep synchronous with the system. The voltage and  power are stable during the fault as shown in Fig. \ref{11}. The above results demonstrate the effectiveness of the proposed current control method.

\subsection{Two-area System}
In this section, the two-area system \cite{kundur1994power} considering the dynamics of ac system is used to test the proposed method. MMC is connected to bus 7 with the double lines. The inductance of single line from bus 7 to 12 is 0.01 $H$, and the inductance of single line from bus 12 to 13 is 0.25 $H$. Three-phase short circuit fault occurs in one of the line 12-13 at 1 $s$. The fault occurs at a position of 0.2 $H$ from bus 12. The fault is cleared at 1.6 $s$.
\begin{figure}[H]
\setlength{\abovecaptionskip}{-0.1cm} 
	\centering
	\includegraphics[scale=0.25] {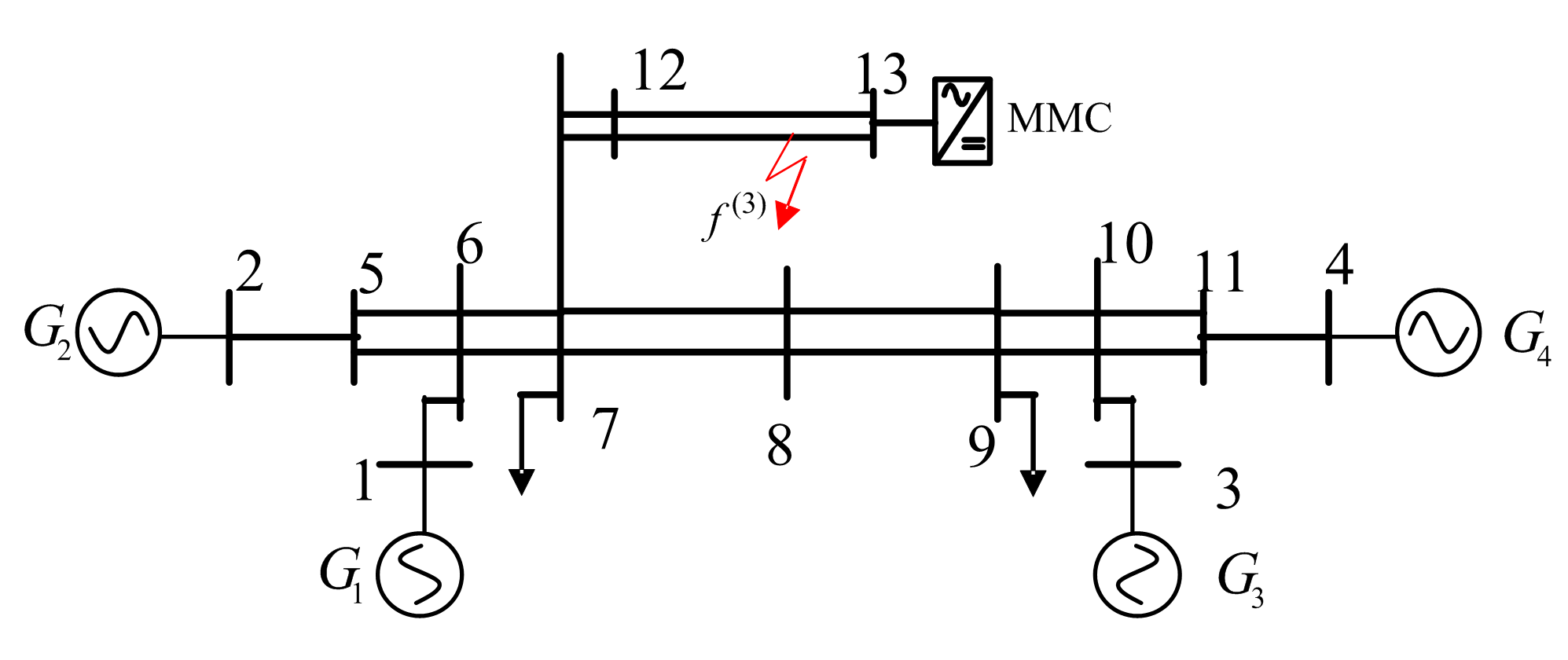}
	\caption{Two-area system with an MMC connected}
	\label{12}
\end{figure}

\subsubsection{TE parameters estimation}
\begin{figure}[H]
\setlength{\abovecaptionskip}{-0.1cm} 
\centering
\includegraphics[scale=0.5]{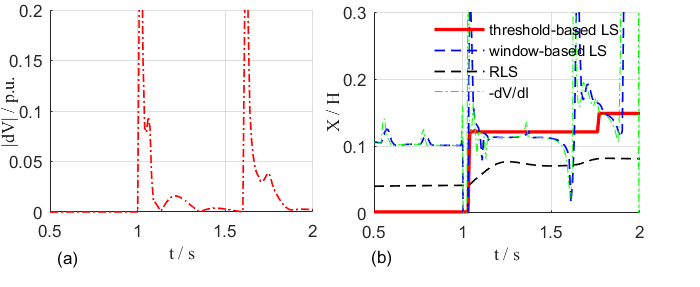}
\caption{TE estimation considering dynamics of ac system. (a) $||dV||$ when subject to a large disturbance; (b) the comparison of TE parameters estimation using different methods}
\label{13}
\end{figure}

the equivalent inductance seen from bus 7 to the system is 0.02 $H$, which can be calculated by the circuit parameters. The equivalent parameters after failure is the same with those in the previous section, which is 0.107 $H$. Fig. \ref{13} shows the voltage variation and inductance estimation results. The equivalent inductance is identified as 0.122 $H$ at 1.04 $s$, and TE potential is $0.287 \, \text{pu}$. Comparing to other methods, the threshold-based LS method can track the inductance faster and more accurately. It is noted that after the fault is clear, the proposed method can still track the post-fault inductance as 0.15 $H$, while the others methods cannot track the inductance even within a longer time. At the same time, the estimation of the post-fault inductance has a large fluctuation when using other methods. That is because when the MMC return stable, the measurements does not change resulting that the system parameters are not observable.

\subsubsection{Output current control of MMC}
\begin{figure}[H]
\setlength{\abovecaptionskip}{-0.1cm}   
\centering
\includegraphics[scale=0.5]{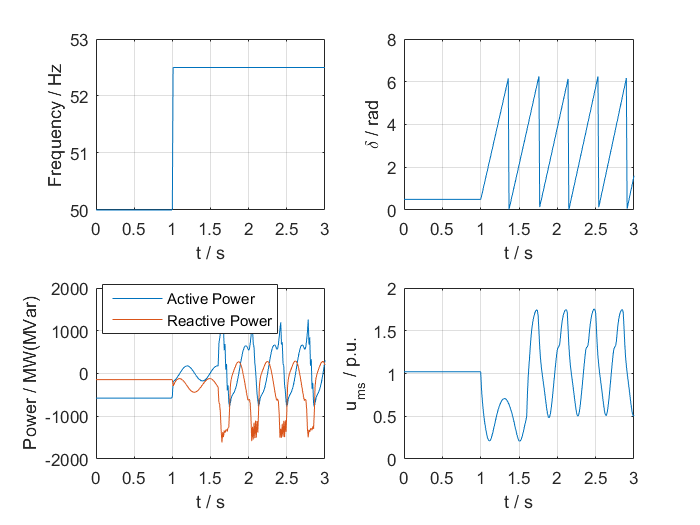}
\caption{{{Simulation curves when $i_{sd}$ = $0.55 \, \text{pu}$. (a) frequency; (b) angle; (c) active/reactive power (d) voltage.}}}
\label{14}
\end{figure}

\begin{figure}[H]
\setlength{\abovecaptionskip}{-0.1cm}   
\centering
\includegraphics[scale=0.5] {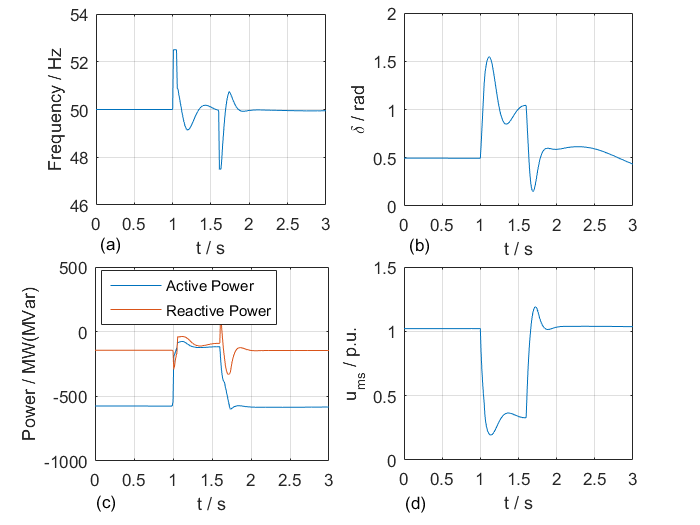}
\caption{{{Simulation curves when $i_{sd}$ = $0.45 \, \text{pu}$. (a) frequency; (b) angle; (c) active/reactive power (d) voltage.}}}
\label{15}
\end{figure}

based on (\ref{e9}), it can be calculated when the output d-axis current of MMC is larger than $0.5\,\text{pu}$, MMC cannot keep in sync with the system.
When $i_{sd}$  = $0.55\, \text{pu}$, it can be seen from Fig. \ref{14} that, the MMC is asynchronous with the system and the voltage and power are unstable. But when $i_{sd}$  = $0.45 \,\text{pu}$ is set, the MMC is synchronous with the system and the voltage and the power output by MMC are stable as shown in Fig. \ref{15}. Above all, the proposed method can still work well when considering the system dynamics.


\section{Conclusion and Future Work}
In this paper, an adaptive control method for synchronization and stability during MMC failure based on the estimation of TE parameters is proposed. This method does not rely on off-line simulation strategy, but can obtain the current instruction to ensure the synchronization and stability of MMC according to the local PMU measurements. To this end, the proposed method is adaptive to various fault conditions. The proposed TE parameter estimation method can quickly identify the grid equivalent parameters after the failure, independent of the initial value, and without the limitation of constant potential on the grid side.

\ifCLASSOPTIONcaptionsoff
  \newpage
\fi

\bibliographystyle{IEEEtran}
\bibliography{IEEEabrv,mybibfile}

\end{document}